# Localization characteristics of two-dimensional quasicrystals consisting of metal nanoparticles

Jian Wen Dong[1,2], Kin Hung Fung[1], C. T. Chan[1,*], and He-Zhou Wang[2]

[1.] Department of Physics, The Hong Kong University of Science and Technology, Hong Kong, China

[2.] State Key Laboratory of Optoelectronic Materials and Technologies, Zhongshan (Sun Yat-Sen) University, Guangzhou 510275, China

Using the eigen-decomposition method, we investigated the plasmonic modes in a two-dimensional quasicrystalline array of metal nanoparticles. Various properties of the plasmonic modes, such as their symmetry, radiation loss and spatial localization are studied. Some collective plasmonic modes are found to be leaky with out-of-plane radiation loss, but some modes can have very high fidelity. To facilitate the analysis on the complex behavior of the collective plasmonic modes, we considered a trajectory map in a three dimension parameter space defined by the frequency of the mode, the eigenvalue that is related to the inverse lifetime of the mode, and the participation ratio of the mode that gives the spatial distribution,. With the help of the trajectory map method, we found that there are modes that are localized in some special ways and the plasmonic mode with highest spatial localization and highly fidelity is found to be a type of anti-phase ring mode. In general, different localized modes have very different radial decay behaviors. There is no special relationship between the fidelity of the modes and their spatial localization.

78.67.Bf, 73.22.Lp, 78.70.-g , 61.44.BrCorresponding author: phchan@ust.hk## I. Introduction

In recent years, there has been growing interest in the plasmon excitation of metal nanoparticles (MNPs) and their aggregates because of the significant improvement in fabrication techniques[1,2] and their plausible applications. These plasmonic materials can support subwavelength phenomena near optical frequencies, and have strong near-field enhancement.[3,4] As such, they can serve as the building blocks for a variety of nanoscale optical devices, such as in biosensors,[5,6] subwavelength waveguides,[7,8] and nanoantenna.[9] Other applications, such as left-handed metamaterials in the optical domain[10] and slow-light effect,[11,12] have also been proposed. The properties of some complementary quasicrystalline plasmonic system (nanohole arrays in metallic films) have been studied in the literature.[13]

Aside from plausible applications, plasmonic systems offer us an interesting platform to explore and discover new physical phenomena that are of academic interest. In this paper, we are going to explore the phenomena associated with the localization of waves in plasmonic quasi-periodic lattices. For electronic bound states, the relationship between disorder and localization is well-understood.[14] The properties of localized and/or critical states in quasicrystalline potentials has also been firmly established.[15] It is also known that localized states of light can be found in disordered[16] or aperiodic photonic crystal systems. In terms of physical mechanism, plasmonic systems are interesting because the underlying physics is somewhere in between electronic and photonic system, but is unique in its own way. For electronic systems, the coupling is due to overlapping wavefunctions and the physics can be captured via short-range models like tight-binding Hamiltonians. The strong plasmonic excitations are intrinsic long-range in coupling, and the fact that all the results are solutions of Maxwell equations means that it is closely related to photonic crystals. However, as the dielectric function is negative for plasmonic systems, the resonances bear resemblance to the bound states in electronic systems, more than to the scattering states in photonic crystals with positive dielectric constants. We first note that the characteristics of electromagnetic (EM) states of 2D arrays of MNPs have already received some attentions in the literature.[11,12,17-20]

In previous studies on the wave transport in electronic and photonic systems, the spatial localization of eigenmodes have been the main focus of theoretical analysis.[21-23] The electronic system has true eigenmodes that are derived from bound states. The localization is thus entirely concerned with the spatial localization of the quasicrystal electronic wavefunctions. For 2D dielectric photonic crystal systems, we are also dealing with true eigenstates, which are derived from scattering states. The localization is still essentially a study of spatial properties of the real space eigenmodes. In electronic or photonic systems, a mode is called localized mode when the electron or photon wave function is confined in a particular region, and decays exponentially in real space, and thus the long distance transport of energy or signal through the media is impeded. However, the modes in MNP systems are not true eigenstates as they may lose energy by coupling to either the dissipative environment or to free photons. In fact, the leaky wave characteristics of quasi-crystalline structures have not been considered until very recently.[24] Once the time dependence is considered, the oscillating frequency can be a complex quantity with an imaginary part, leading to a finite life time. When we consider the localization of plasmonic modes, we should pay attention to their spatial profile as well as their fidelity (quality factor). The fidelity of the state enters as an extra degree of complexity, and an extra dimension should be added in order to describe the fidelity or the radiation loss of the localized modes. Thus, we are dealing with the localization properties of "leaky" states with long range couplings in a quasi-crystalline lattice. The physics is anticipated to be richer and more complex than those in electronic and photonic crystal systems. Up to now, the Anderson localization of eigenmodes in random MNP arrays has been studies.[17-19] In this paper, we are concerned with the localization of the plasmonic modes on a 2D QC lattices. The major difference between quasiperiodic and random MNP is that the latter does not possess the type of high symmetry local structures (such as the rings in 12-fold lattice) that can anchor strongly localized states. Random structures also do not have the deterministic "recurrence" of local structures that leads to self-similar mode profiles.

In this paper, we will focus on the analysis of the out-of-plane plasmonic modes in 2D quasicrystalline array of MNPs. We found rich and complex localization characteristics of the plasmonic modes, such as their rotational symmetry, radiation loss, and spatial localization. All the modes are leaky, and some plasmonic modes have strong out-of-plane radiation loss, while there is a set of modes that are weakly leaky and have high quality factors. To study the complex EM localization characteristics of the plasmonic modes, we employ a trajectory map which enables us to find the modes that are localized in different ways. In particular, an anti-phase ring (APR) mode is found to be the highest spatially localized and has high fidelity; while an in-phase ring (IPR) mode is highly spatially localized but has high radiation loss. We also observed that different localized modes have very different radial decay behaviors. Some of them are rather specific for the quasi-crystalline plasmonic lattices, and very different from previous electronic or photonic systems. The physical mechanisms of the decay behaviors in the APR and IPR modes are discussed.

This paper is organized as follows. In Section II, the geometry and the material parameters of the 2D QC, as well as the eigendecomposition (ED) method are described. In Section III, the complex spatio-temporal characteristics of the plasmonic modes are discussed. The conclusions are presented in Section IV.

## II. Physical system and method of calculation

In this work, we study the properties of the collective plasmonic modes in the 2D QC and we choose a QC with 12-fold symmetry generated by the generalized dual method[25] as prototype. The 12-fold dodecagonal pattern is tiled by three basic cells (a square and two rhombi with 30° and 60° angles). The plasmonic system is then constructed by placing metal nanoparticles with circular cross sections in the sites of a 2D dodecagonal lattice. The structure of the plasmonic lattice is shown in Fig. 1(a). Theoretical and experimental investigation of the band gap properties of dielectric photonic quasicrystals of the same symmetry has been considered previously.[22] The plasmonic structure is in an air background ($\varepsilon_h = 1$). The radii of MNPs are $r_0$, which is 25 nm. The distance between two nearest particles is $a_{min}$, which is 75 nm. We restrict ourselves to the situations when the MNPs are not too close together, so that $a_{min} \geq 3r_0$, and at such a distance, the system can be treated very accurately with the dipole approximation. For comparison, the structure of the square lattice is also shown in Fig. 1(b), whose lattice constant is $a_{latt} = a_{min} = 75 nm$.

To describe the EM resonances of the QC array, a good approach is the ED method[26-28] (sometimes called the spectral decomposition method). Bergman *et al.*[26] proposed such a method that considers all multipolar responses of the studied system. Markel *et al.*[27] developed a simplified version that considers only the dipolar response. The dipolar ED method was applied successfully to systems such as disordered arrays[17], periodic MNP,[12] and circular arrays.[20] The key advantage of this approach is that it does not require either numerical complex root searching[29] or root approximation[30] in the complex number plane. In addition, the method is efficient. It readily gives the resonant spectrum, quality factor, and participation ratio simultaneously, and these quantities are directly related to responses of the system. Moreover, it is applicable to an open system with radiation and absorption losses.

Let us consider an external driving electric field of the form $\mathbf{E}_m^{ext} e^{-i\omega t}$, which acts on a lattice of $N$ particles located at points $\mathbf{r}_1, \mathbf{r}_2, ..., \mathbf{r}_N$. Each particle has the same dynamic dipole polarizability $\alpha$, which is given by[31]

$$\alpha(\omega) = i\frac{3c_h^3}{2\omega^3} a_1(\omega).  \tag{1}$$

Here,

$$a_1(\omega) = \frac{q\psi_1(qx_0)\psi_1'(x_0) - \psi_1(x_0)\psi_1'(qx_0)}{q\psi_1(qx_0)\xi_1'(x_0) - \xi_1(x_0)\psi_1'(qx_0)}  \tag{2}$$

is the "$\ell = 1$" electric term of the Mie's coefficients,[32] $\psi_1$ and $\xi_1$ are the Riccati-Bessel functions, $x_0 = \omega r_0/c_h$ ($c_h$ is the speed of light in the background medium), and $q = \sqrt{\varepsilon(\omega)/\varepsilon_h}$. The dielectric function $\varepsilon(\omega)$ of the MNPs is chosen to be the Drude's form,

$$\varepsilon(\omega) = 1 - \frac{\omega_p^2}{\omega(\omega + i\gamma)},  \tag{3}$$

where $\omega$ is the angular frequency, $\omega_p$ is the plasma frequency (which is 6.18 eV in this paper), and $\gamma$ is the electron scattering rate. The coupled dipole equations can be written as

$$\mathbf{p}_m = \alpha\left[\mathbf{E}_m^{ext} + \sum_{m'\neq m} \overleftrightarrow{\mathbf{W}}(\mathbf{R}_m - \mathbf{R}_{m'})\mathbf{p}_{m'}\right],  \tag{4}$$

where $\mathbf{p}_m \equiv (\mathbf{p}_{m,//}, p_{m,z})^T$ is the $m$-th dipole moment, and the dynamic Green's function is,[27]

$$W_{uv}(\mathbf{r}) = k_0^3[A(k_0 r)\delta_{uv} + B(k_0 r)\frac{r_u r_v}{r^2}],  \tag{5}$$

$$A(x) = (x^{-1} + ix^{-2} - x^{-3})e^{ix},  \tag{6}$$

$$B(x) = (-x^{-1} - 3ix^{-2} + 3x^{-3})e^{ix},  \tag{7}$$

where $k_0 = \omega/c_h$ and $u, v = 1, 2, 3$ are component indices in Cartesian coordinates. Equation (4) can be written in a compact form as

$$\sum_{m'=1}^{N}\sum_{v=1}^{3} M_{mm'uv} p_{m'v} = E_{mu}^{ext}, \tag{8}$$

or $\mathbf{Mp} = \mathbf{E}$ in a matrix form. Here, $\mathbf{M}$ is a $3N \times 3N$ matrix, and can be divided into two parts $\mathbf{M} = \alpha^{-1}\mathbf{I} - \mathbf{G}$, where $\alpha^{-1}\mathbf{I}$ describes material properties, and $\mathbf{G}$ describes a lattice Green's function which is determined by the specific geometrical structure of the lattice; and $\mathbf{p}$ is a $3N$-dimensional column vector. To analyze the resonances of a cluster of dipoles, we consider the following eigenvalue problem:

$$\mathbf{M}|\hat{\mathbf{p}}\rangle = \lambda |\hat{\mathbf{p}}\rangle, \tag{9}$$

where $\lambda$ and $|\hat{\mathbf{p}}\rangle \equiv (...,\hat{\mathbf{p}}_{m-1}, \hat{\mathbf{p}}_m, \hat{\mathbf{p}}_{m+1},...)^T$ are the complex eigenvalues and the complex eigenvectors of the interaction matrix $\mathbf{M}$. The eigen-polarizability, which describes the response of a particular mode, is defined as[11]

$$\alpha_{eig} = \frac{1}{\lambda}. \tag{10}$$

This quantity can be interpreted as the collective response function of the whole system for an external electric field pattern that is proportional to the corresponding eigenvector. Both $\lambda$ and $\alpha_{eig}$ are useful quantities for analyzing the intrinsic normal modes of the system.

In general, there are $3N$ eigenmodes for a particular system. In this paper, we focus on the out-of-plane modes ($\hat{\mathbf{p}}_{m,//} = \mathbf{0}$ and $\hat{p}_{m,z} = \hat{p}_m \neq 0$). The dimension of the interaction matrix $\mathbf{M}$ can be reduced to $N \times N$, and the number of eigenmodes at each frequency becomes $N$. In spite of this simplification, it is difficult to study the plasmonic modes analytically. In fact, plasmonic modes of 2D aperiodic structures cannot be obtained analytically when a large number of particles is involved.[17] Thus, in solving the coupled-dipole equation [Eq. (9)], we use standard matrix diagonalization routines (LAPACK) for the diagonalization of the interaction matrix, $\mathbf{M}$.

As the matrix $\mathbf{M}$ is frequency dependent, we obtain $N$ eigenvalues and $N$ eigenvectors for each frequency. The eigenvectors for a particular eigenvalue at a given frequency give the spatial distribution of the plasmonic modes, while the inverses of the eigenvalues give the response of the modes to the excitation at that particular frequency. As the amount of information contained in the eigenvectors are overwhelming, it would be useful to consider some characteristic numbers that tell us something useful about the spatial properties of the mode. The participation ratio (PR), defined as,

$$P_n = \left(\sum_{m=1}^{N}\left|\hat{\mathbf{p}}_m^{(n)}\right|^2\right)^2 \bigg/ \left(\sum_{m=1}^{N}\left|\hat{\mathbf{p}}_m^{(n)}\right|^4\right), \tag{12}$$

is a good indicator of the spatial localization properties of the modes. We use the symbol $P_n$ to refer to the $n$-th plasmonic eigenmode. The value of $P_n$ varies significantly between spatially extended and localized modes. For example, if the intensity of an extended eigenfunction at each site is equal, we have $P_n = N$. If the mode is completely localized at a single site, we have $P_n = 1$. In general, a mode can be regarded as localized when $P_n \ll N$.

## III. Numerical results and discussions

In this section, we will study the spatio-temporal characteristics of the plasmonic eigenmodes in the 12-fold QC. The results on the rotational symmetry, Fourier analysis, and spatial localization of the plasmonic eigenmodes are shown in Sec. III A to III C. At the end of this section, we will focus on the spatial decay behaviors of some highly localized modes. In this paper, we focus on the effect of the structure of the quasi-periodic arrangement on the localization character of the plasmonic system, and for that purpose, we set $\gamma = 0$ (no metal absorption) to avoid extra complications. Otherwise, it is difficult to delineate whether the spatial decay is due to absorption or due to the scattering effect of the quasi-periodic arrangement. In addition, we consider MNP lattices with up to 36000 particles, which correspond to a lattice size of about 26 μm in one dimension. In view of fabrication limitations and the intrinsic dissipation in plasmonic systems, larger lattices would probably be of mathematical interest only.

### A. Rotational symmetry

Before we show the numerical results, let us use symmetry to classify the eigenmodes. Since the QC that we consider has a 12-fold discrete rotational symmetry (represented by the cyclic group, $C_{12}$), we can simplify the eigenvalue problem by dividing the lattice points into $M+1$ sets, where $M \equiv (N-1)/12$, so that each set (except the one which contains only the central point located at the origin) contains 12 symmetric lattice points that lie on a single ring. We use the label $s$ to indicate the $s$ th set of lattice points (counted from the ring with smallest radius) and the label $t$ to indicate the $t$ th lattice point in each set (counted from the $x$-axis in anti-clockwise direction). For the out-of-plane modes, the dipole moment at each point (except the origin) can be denoted as $\hat{p}_{s,t}$, where $s = 1, 2, \ldots, M$ and $t = 1, 2, 3, \ldots, 12$ while the only one at the origin is denoted as $\hat{p}_{0,0}$. With the help of the group representation theory, we know that the eigenmodes take the form,

$$\hat{p}_{s,t}^{(j,k)} = c_s^{(j,k)} e^{i2\pi jt/12} \text{ for } s \neq 0, \tag{13}$$

where $j = 1, 2, 3, \ldots, 12$ and $k = 1, 2, 3, \ldots, M$. Here, we replace the mode index $n$ by a pair of $j$ and $k$. In Eq. (13), we see that, for each eigenmode, the dipole moments on the lattice points in each set, $s$, must have the same magnitude. Substituting Eq. (13) into Eq. (9) for the out-of-plane modes, we can reduce the eigenvalue problem to one $M \times M$ matrix equation for each of $j = 1, 2, 3, \ldots, 11$ and one $(M+1) \times (M+1)$ matrix equation for $j = 12$. The in-phase ($j = 12$) modes and anti-phase ($j = 6$) modes are, in general, non-degenerate while, due to the mirror symmetry of the QC, the modes indexed by $j = 1, 2, 3, 4,$ and 5 are degenerate with that indexed by $j = 11, 10, 9, 8,$ and 7, respectively. Except that there may be accidental degeneracy at a particular frequency, the above analysis is independent of frequency. In the following calculation, we usually employ a circular domain to satisfy the requirement of rotational symmetry analysis. Another advantage to use circular domains is that the application of rotation symmetry can block-diagonalize the matrix and thus allow us to go to much larger samples. Since the main interest of this work is focused on a subset of eigenmodes that are highly localized, the boundary has no effect on them as long as the lattice size is large enough.

**B. Radiation (fidelity) properties**

In a periodic 2D array of MNPs, the plasmonic modes can easily be classified either as radiating (leaky) or not simply by examining the dispersion diagram, because each mode has a well-defined Bloch index.[11,12] For those modes that have Bloch $k$ vectors with magnitudes less than $\frac{\omega}{c_h}$, the mode can couple with free photons and the life time will be finite due to radiation loss. If $k > \frac{\omega}{c_h}$, the mode becomes a guided mode (confined in the 2D plane) that has no radiation loss. However, such an analysis is not suitable for the QC systems due to the absence of translational symmetry. One can consider the fidelity of the mode by examining the imaginary part of the eigen-polarizability, $\text{Im}(\alpha_{eig})$. The quality factor of a particular eigenmode is high when the magnitude of $\text{Im}(\alpha_{eig})$ is large (or alternatively, the $|\lambda|$ is small) at its resonant frequency. In Fig. 2(a), we plot $\text{Im}(\alpha_{eig})$ as a spectral function of frequency by diagonalizing the interaction matrix **M** in the 12-fold QC with $N = 1513$. When we diagonalize the matrix **M** using standard numerical routines in LAPACK, we get $N$ eigenvalues for each frequency, and there is no obvious ways to connect the eigenvalues together from one frequency to another. To help identify modes that exist at different frequency regimes, we use different colors for different frequency regions. There are three regions of resonant bands from 3 to 3.8 eV. The black (near 3.25 eV), yellow/red (above 3.45 eV), and pink (below 3.2 eV) bands are denoted as band I, II, and III, respectively. It can be seen that the $\text{Im}(\alpha_{eig})$ in band I is much higher than those in band II and III. It shows that the quality factor of band I is higher.

In order to understand why the modes in band I has particularly high quality factor, we calculate the Fourier spectra of some eigenmodes in the QC, and for the ease of numerical computation, the eigenmodes of the QCs are calculated with square-shape boundaries.[33] Figure 3 shows the Fourier spectra of three interesting modes which have resonant frequencies that are located in different resonant bands. In Fig. 3(a), we show a mode with high quality factor with its resonant frequency inside Band I [the black region in Fig. 2(a)]. This is an anti-phase ($j = 6$) mode. The excited dipole moments of this mode are strongly confined on the central ring of the QC. The magnitude of the dipole is exactly the same in every particle within the central ring, but the direction of adjacent dipoles are opposite in sign. The dipole moments are shown pictorially in the insets. We shall call this mode an APR mode for obvious reasons. The Fourier transform of the spatial dipole moment distribution shows that the APR mode has a broad bandwidth in momentum space, which is expected for a spatially localized mode. The peaks of the distribution are outside the light line. This is also understandable from the spatial distribution. A chain of alternate up and down dipoles spatially close to each other should have small projections on small $k$-vectors when we do Fourier transform. Since most of the $k$ components satisfy $k > \dfrac{\omega}{c_h}$, this mode radiates very weakly to the third dimension and hence the mode has high quality factor. We shall call this type of modes weakly-radiating modes. All the high quality modes in Band I are weakly-radiating modes and these modes are characterized by eigenvectors that have an equal number of up and down dipoles of equal magnitudes that are positioned very close to each other.

Figure 3(b) shows the most spatially localized mode within the frequency range in Band II. The excited dipole moments are again strongly localized on the central ring and the central particle of the QC lattice. The magnitudes of the dipole moments on the ring are the same and the directions between two nearest particles are the same (in-phase), as shown in the insets. We call this mode an IPR mode. The Fourier transform of the dipole moment distribution of this mode is peaked inside the light line. This is rather reasonable since a chain of equal dipoles spatially close to each other should have big projections on small $k$-vectors and small projections on large $k$-vectors. Since most of the $k$ components satisfy the condition $k < \frac{\omega}{c_h}$, this is a leaky mode which has strong out-of-plane radiation. The quality factor is thus smaller than the modes in Band I.

We now come to another type of modes which is picked from Band II in the peak of the yellow region [see Fig. 2(a)]. The dipole moments are shown in the insets of Fig. 3(c). The dipoles form a ring-like pattern, and as the ring has a radius that is very big compared with inter-particle distances, we call this type of modes "large-ring" (LR) modes. The ring-like pattern has patches of localized dipoles, and the phase of the moments within each patch is the same, but the phases between adjacent patches are opposite. The Fourier transform of the dipoles shows that the eigenvectors have $k$-components that are essentially within the light line, and for this reason, these leaky modes couple well with free space photons and their fidelities are not high. We also found that the modes in Band III have $k$-components that are strongly peaked inside the light cone. These modes are also strongly radiating and have low quality factors.

We show in Fig. 2(b) the eigen-polarizability spectrum of a square array for comparison. In order for a meaningful comparison, the square array has exactly the same number of particles as the QC. We immediately see some correspondences between the modes and we label these modes with the same color. In the square array of this size, the Band I modes have frequency components essentially outside the light cone. As the size of the square array increases, the Fourier transformed spectra will peak at one single $k$ which is outside the light cone, indicating that the mode becomes a truly guided Bloch mode which is periodic on the 2D plane, and the value of $\text{Im}(\alpha_{eig})$ will increase without bound for these Band I modes in the square array. In contrast, the $\text{Im}(\alpha_{eig})$ for the QC will not increase indefinitely even if the size of the QC increases. These modes remain weakly radiating as there are always some (albeit very small) components inside the light cone. For the Band II modes in the square array, they have projections inside the light line peaked near $k=0$ even if the size of the array increases. They are leaky resonances. Both the guided modes and leaky resonance have been identified in the dispersion diagram of the infinite square array of MNPs.[12]

**C. Localized modes**

In a perfect periodic infinite system, all modes are Bloch modes, which form a truly continuous spectrum and can be labeled by a continuous $k$ vector in the first Brillouin zone of the lattice. If we now break the translational symmetry of a sample, the modes are not Bloch modes and can be either localized or delocalized. In the following, we will introduce a trajectory map to analyze the modes.

The eigenmodes are displayed using a trajectory map defined in a three-dimensional (3D) parameter space ($\omega$, $|\lambda|$, $P_n$), which can give information about the eigenvalue (related to the inverse of lifetime, and hence temporal properties) and the participation ratio (related to the spatial properties of the modes). The 3D trajectory map of a 12-fold QC with 1513 particles is shown in Figs. 4(a) and 4(b), which shows that the extremely dense points in Fig. 2 can be connected by continuous trajectories. To show the trajectory map in a clearer way, we plot its 2D projection in the $|\lambda|$-$P_n$ plane [see Fig. 5(a) in a larger scale, and Figs. 5(c), 5(e) in expanded views with small values of $P_n$]. From the 2D projections (top view), we see that $P_n$ varies from 12 to ~1000, indicating that the QC has both highly localized and highly delocalized eigenmodes. For comparison, the result for a larger 12-fold QC with 3529 particles is also shown in Figs. 5(b), 5(d), and 5(f). In the larger scale, the point distributions in Figs. 5(a) and 5(b) are very similar except that the range of $P_n$ are stretched to approximately twice the value by increasing the size of the QC. This indicates that there are many delocalized modes whose profiles will expand when the sample size increases. However, there are trajectories that are nearly unchanged regardless of the size of the QC and these can be found on the bottom-left region of the 2D projections which are for small values of $P_n$ and $|\lambda|$. The arrows in Figs. 5(c)-(f) highlight the trajectories of three localized modes. It is clear that the $P_n$ of the APR mode remains the same independent of the size of the sample, which is a signature of localized modes. The same can be observed in the IPR modes. For the LR mode, the appearance of the trajectories ($P_n$ is between 200 and 250) are almost invariant with the increase of the size of the QC. Meanwhile, the $P_n$ of the modes in this stable region are much smaller

than the number of particles. For example, the $P_n$ of two highest localized modes (the APR and IPR mode) are about 12 and 13, respectively, in the QC with thousands of particles. In general, a mode can be regarded as localized if $P_n \ll N$. Therefore, the modes in this region are indeed spatially localized.

Although the modes are spatially localized, they do have distinct properties. In Figs. 5(c)-(f), it is clear that the minimum of $|\lambda|$ reach different values for different trajectories. For the APR mode, the minimum of $|\lambda|$ approaches to a near-zero value due to its high fidelity. This is very different from the case of the IPR (or LR) modes, of which the minimum is about 0.04 (or 0.01), and the lower fidelity is due to their stronger out-of-plane radiation. Generally speaking, we do not find any special relationship between spatial localization and fidelity of the mode. More evidences can be found in the 3D trajectory map. In Figs. 4(a) and 4(b), we see that modes with small values of $P_n$ (spatially localized mode) can either be weakly-radiating (Band I modes, black) or strongly-radiating (Band II, yellow). It is also the case for those modes with large value of $P_n$. In addition, the appearances of the trajectories are also different from each other. For some localized modes, the $P_n$ varies quickly with the frequency, indicating that the mode patterns will change a lot between on- and off-resonance. On the other hand, there are also modes in which $P_n$ remains essentially the same regardless of the frequency, indicating that the mode patterns remain invariant in the studied frequency region.

To show that the existences of these localized modes are the special properties for the QC that we consider, we compare the trajectory maps with that of square lattices. Results for square lattices of different sizes are shown in Figs. 4(c), 5(g) and 5(h). There are no trajectories in the region with small value of $P_n$ and $|\lambda|$. There are also no stable regions in the trajectory map in which the values remain invariant. This is because the eigenmodes for finite-size square lattices are close to delocalized Bloch modes whose PR value is not small and scales with the lattice size. As the particles of the lattice increase from 1513 to 3613, the eigenmodes become more extended so that the range of $P_n$ are expanded to approximately twice the value. We note that the modes, of which the $P_n$ is about 250 (500) in Fig. 5(g) [Fig. 5(h)], are the boundary modes. The intensity of dipole moments of such modes is confined along the boundary of the finite square lattice.

**D. Decay behaviors of localized modes**

We have also calculated the average intensity $I(r)$ of the mode functions as a function of distance $r$ from the center of the sample so as to study the localization behaviors of the plasmonic eigenmodes. The average intensity is calculated by integrating the dipole intensity $\left|\hat{p}_m^{(n)}(\mathbf{r}')\right|^2$ in an annular circle defined by $r - \frac{\delta r}{2} < |r'| < r + \frac{\delta r}{2}$ and dividing the result by the area, $2\pi r \delta r$, with $\delta r = 6R_0$, where $R_0$ is the radius of the central ring in the QC. We will focus on the regional behavior of the spatial distribution within a few microns. Within such a range, we will see that the envelope of $I(r)$ exhibits both power-law and exponential behaviors, and the character is highly dependent on the mode profile.

For most of the modes, the spatial profiles change as the size of the system changes and they change in a manner that seems to defy a simple classification. However, there are some highly localized modes in which the spatial profiles are essentially invariant as the size of the QC lattice grows. Figure 6 shows three such highly localized eigenmodes, corresponding to the modes shown in Figs. 3 (a)-(c). The envelopes of $I(r)$ for different sizes of the QC arrays are shown in the figure when these modes are at resonance (i.e. we pick the frequency when $|\lambda|$ is minimum). In Fig. 6(a), the envelope of $I(r)$ of the APR mode is plotted on a log-log scale. For this mode, the intensity is strongly concentrated on the first ring [as shown in Fig. 3(a)] and then drops abruptly to a small value with a power-law tail outside the ring. Figure 6(a) shows that the tails of the $I(r)$ for different QC sizes fall on the same line, which can be fitted approximately with a power-law decay of $I(r) \propto 1/r$ (see the dash fitting line). When we consider the Fourier spectrum of this APR mode, we already see that most of the $k$-components lie outside the light line. In other words, this mode loses very little energy to the out-of-plane directions. But as most of the dipole intensity are localized on a ring, it cannot be a true eigenstate (see the discussions in Ref. [20]), and it must radiate some energy out. So, the small amount of energy leaked out of the inner ring is confined strongly to the in-plane directions. This is why the tail of $I(r)$ follows approximately the power-law decay of $I(r) \propto 1/r$ as required by energy conservation.

However, such a power-law behavior is not universally true for all the eigenmodes. In particular, the radiating modes have their own localization characteristics. The governing fact is that when the mode quality is low, there will be significant radiation loss to the third dimension in free space. The decay characteristics become more complex than that of the APR mode. We first look at the IPR mode. The envelopes of $I(r)$ for the IPR mode in different sizes of QCs are shown in Fig. 6(b) in the log-linear scale. There are two regimes of decay behaviors: $I(r)$ is a straight line at the region of $r<11R_0$ which manifests some form of exponential decay near the center of the QC, going over to a long-range non-exponential decay when $r>11R_0$. The long range part can be fitted reasonably well with a function $1/r^4$ and so the long range decay has a quasi-power-law behavior. We use the term "quasi-" here because the decay period is not long enough for an unambiguous assignment of power-law decay. This is limited by the self-similar properties of the QC, which are to be discussed later.

We now take a closer look at the short-range decay of the IPR mode, which is characterized by dipoles of equal magnitude and phase on one ring. A ring mode with even parity can radiate strongly.[20] From the Fourier analysis in Fig. 3(b), it has strong out-of-plane radiations as the Fourier components of the mode peak within the light cone. When energy from the inner ring is coupled to the exterior rings, the first ring radiates some energy to out-of-plane directions so that only a small percentage reaches the second ring (we can group the particles in concentric rings about the center of the QC, see Sec. III A for discussion). When the second ring is excited, the dipoles are also in-phase within the ring, so it will also radiate to out-of-plane directions, leaving a small percentage to excite the third ring. As a certain percentage of the energy is depleted to the third dimension as energy is passed from the one ring to another, we observe an exponential decay of the mode profile and the dipole intensity drops to a very small number at $r \sim 11R_0$. We can view this short-range behavior as dominated by a leaky mode on the 2D plane. There is another mechanism in which energy can pass from the inner ring to the outside on the plane. Each dipole is radiating energy to the outside, irrespective of whether there are other particles in between. Each dipole should contribute a $|E|^2 \sim 1/r^2$ to the far field on the plane. However, the IPR mode has the property that the phases of the dipoles have oscillations in the first few rings in the radial direction (see Fig. 7), and the interferences of fields due to various rings lead to the cancellation of the lowest order term in the far field, leaving a higher order tail of $\sim 1/r^4$. As the QC size increases, the power-law tail emerges as the short-range exponential has already decayed to a negligibly small number.

We note in Fig. 6(a) that there is another small resonant excitation at $r \sim 48R_0$ in the large-sized QC structure with 10533 particles. According to Conway's theorem, when the QC sample size is sufficiently large, similar local geometries can appear in the sample quasi-periodically. This leads to the self-similar modes in a sufficiently large QC structure. In fact, the twelve-particle ring pattern will emerge when the radius of the sample reaches $r \sim 48R_0$. When these rings appear, the APR mode that is localized at the inner ring can also be excited at these rings. As a consequence, similar APR patterns appear 12 times at $r \sim 48R_0$ at the 12 rings surrounding the center. This is also true for the IPR mode. To further demonstrate the effect, the IPR mode of a larger QC with 35269 particles is calculated, and the mode pattern is shown in Fig. 8 (in a log scale for easier visual identification). The self-similarity of the IPR mode is obvious.

In Fig. 6(c), we also plot the envelopes of $I(r)$ of the LR mode for different sizes of the QC. It can be seen that the $I(r)$ reaches maximum value at $r \sim 12R_0$ for its large ring-like pattern [see insets of Fig. 3(c)]. The envelopes of $I(r)$ manifest two different forms of exponential decay near the maximum value. This is due to the in-phase excited dipole moments in each patch of the LR mode pattern, which is similar to the IPR mode. As the QC size increases, the structures that are similar to those at $r \sim 12R_0$ will emerge quasi-periodically at $r \sim 35R_0$. Thus, the ring-like patterns appear again, so that another small peak appears in the envelopes of $I(r)$ when the particles of the QC is large than 5677.

## IV. Conclusion

In summary, we have systematically studied the collective plasmonic modes of MNPs in the 12-fold quasicrystalline lattice within the dynamic dipole approximation. Rich and complex spatio-temporal localization characteristics arising from the quasi-periodic arrangment was found. The symmetry of the system implies that the plasmonic modes can be classified in term of the rotational symmetry. Some plasmonic modes are found to be of high quality factor, while some have strong radiation loss. It can be well understood by their spatial mode distributions, as well as their Fourier transforms. The trajectory map in a 3D parameter space of frequency, participation ratio of the mode, and the eigenvalue of the mode, is employed to analyze the complex behaviors of the plasmonic modes. With the help of such a map, we found a rich variety of spatial localization characteristics. In particular, the APR mode has the highest spatially localized as well as the highest fidelity; while the IPR mode is highly localized and has high out-of-plane radiation loss. There are no specific relations between spatial localization and fidelity of the modes. We also found that the radial decay behaviors of the plasmonic modes are highly dependent on mode patterns. For the APR mode, the power-law decay of $1/r$ is found due to the strong energy confinement in the 2D plane. For the IPR mode, there is the quasi-power-law decay after the exponential decay due to the complex radiation loss in the out-of-plane directions. The expected self-similarity of the plasmonic modes is also found if the sample is big enough.

## ACKNOWLEDGMENTS

We would like to thank Prof. Z. Q. Zhang, Dr. Jack Ng, Dr. Yun Lai, and Dr. Zi-Lan Deng for useful discussions. This work was supported in part by the Central Allocation Grant from the Hong Kong RGC through HKUST 3/06C. J. W. D. and H. Z. W. are also supported by the National Natural Science Foundation of China (10804131, 10874250, 10674183), National 973 (2004CB719804), and Natural Science Foundation of Sun Yat-Sen University (2007300003171914).

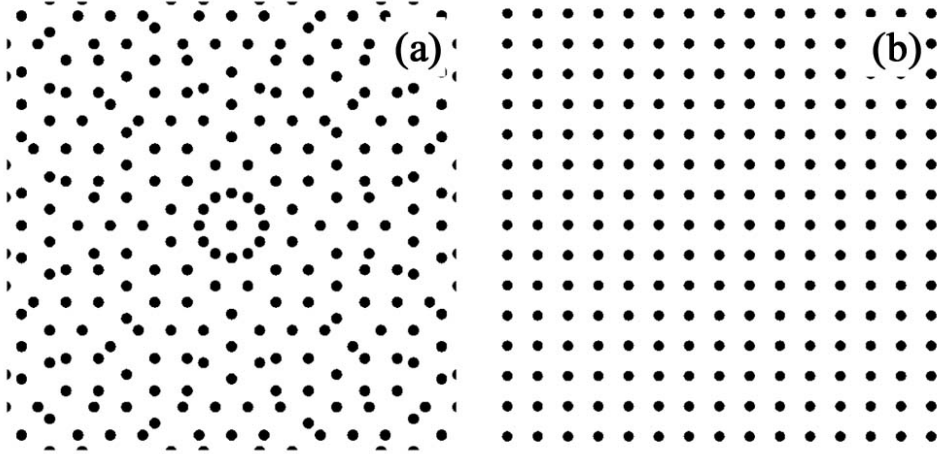

**FIG. 1: Schematic diagrams of the 2D MNP arrays: (a) 12-fold QC and (b) square lattice. In the 12-fold QC case, a circular domain is always employed to satisfy the requirement of rotational symmetry, which can speed up the calculation and allow us to go to much larger samples.**

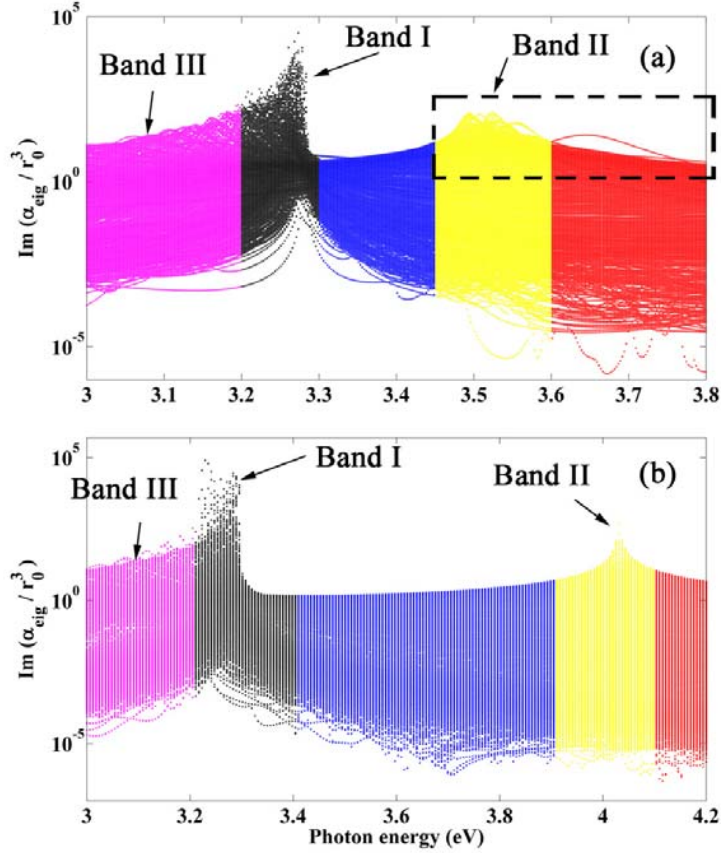

**FIG. 2:** (Color online) Imaginary parts of eigen-polarizabilities $\operatorname{Im}(\alpha_{eig}/r_0^3)$ versus frequency for the out-of-plane plasmonic modes of the (a) 12-fold QC and (b) square lattice. The spectra are in log-linear scale, and different colors mark different frequency regions. $\alpha_{eig}$ is defined as the reciprocal of the eigenvalue $\lambda$ (Eq. 10). Larger $\operatorname{Im}(\alpha_{eig}/r_0^3)$ means the mode is higher fidelity (quality factor) and longer lifetime. The Band I modes are of high fidelity and they are weakly radiating; while the Band II and III modes are radiating. The material parameters for the QC and square lattice are $r_0 = 25\,\text{nm}$, $\omega_p = 6.18\,\text{eV}$, $\omega_p = 6.18\,\text{eV}$, $\gamma = 0$, $\varepsilon_h = 1.0$, $a_{\min} = a_{latt} = 75\,\text{nm}$, and $N_{QC} = N_{sq} = 1513$.

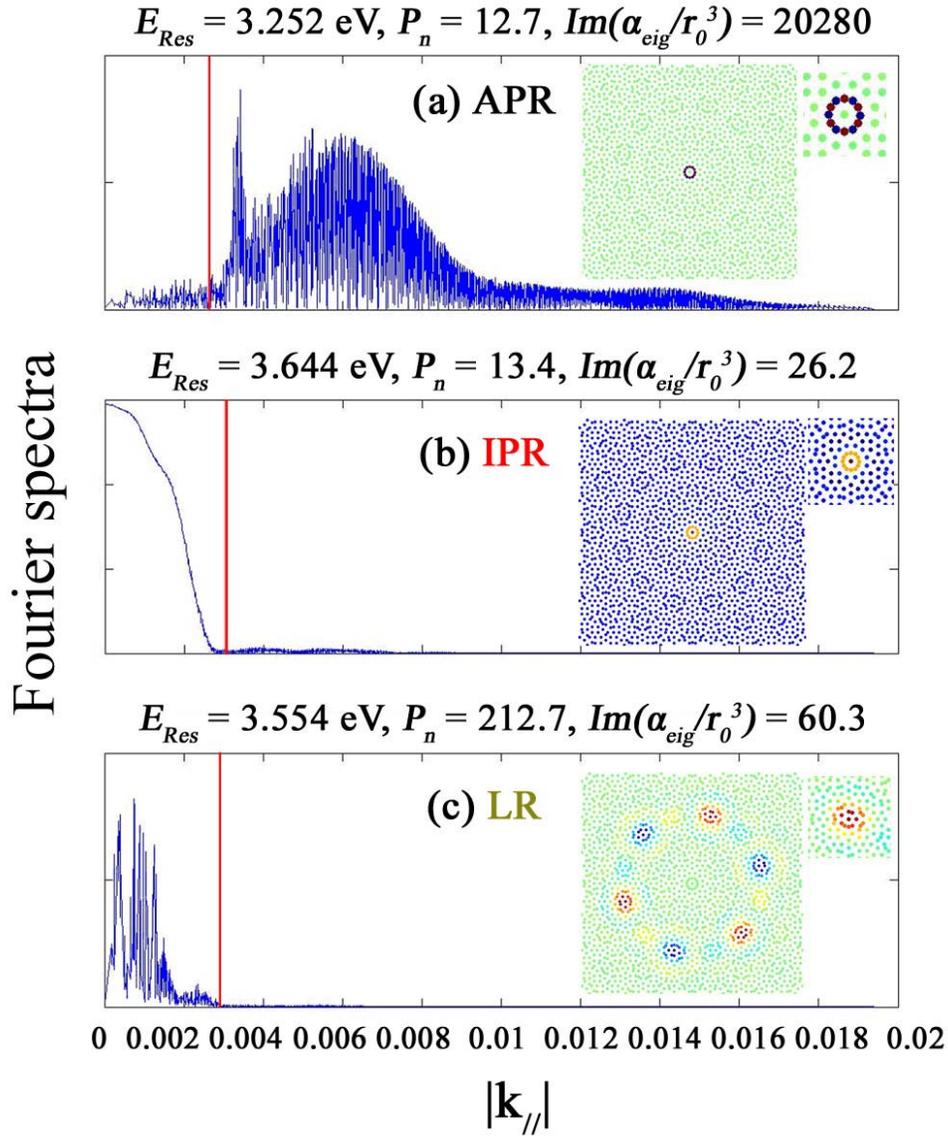

**FIG. 3:** (Color online) Fourier spectra of the (a) APR, (b) IPR, and (c) LR modes. Insets are their mode profiles. The particle number is 1961 in order to form the square boundary of the finite QC for the ease of the calculation. Other structural and material parameters are the same as Fig. 2.

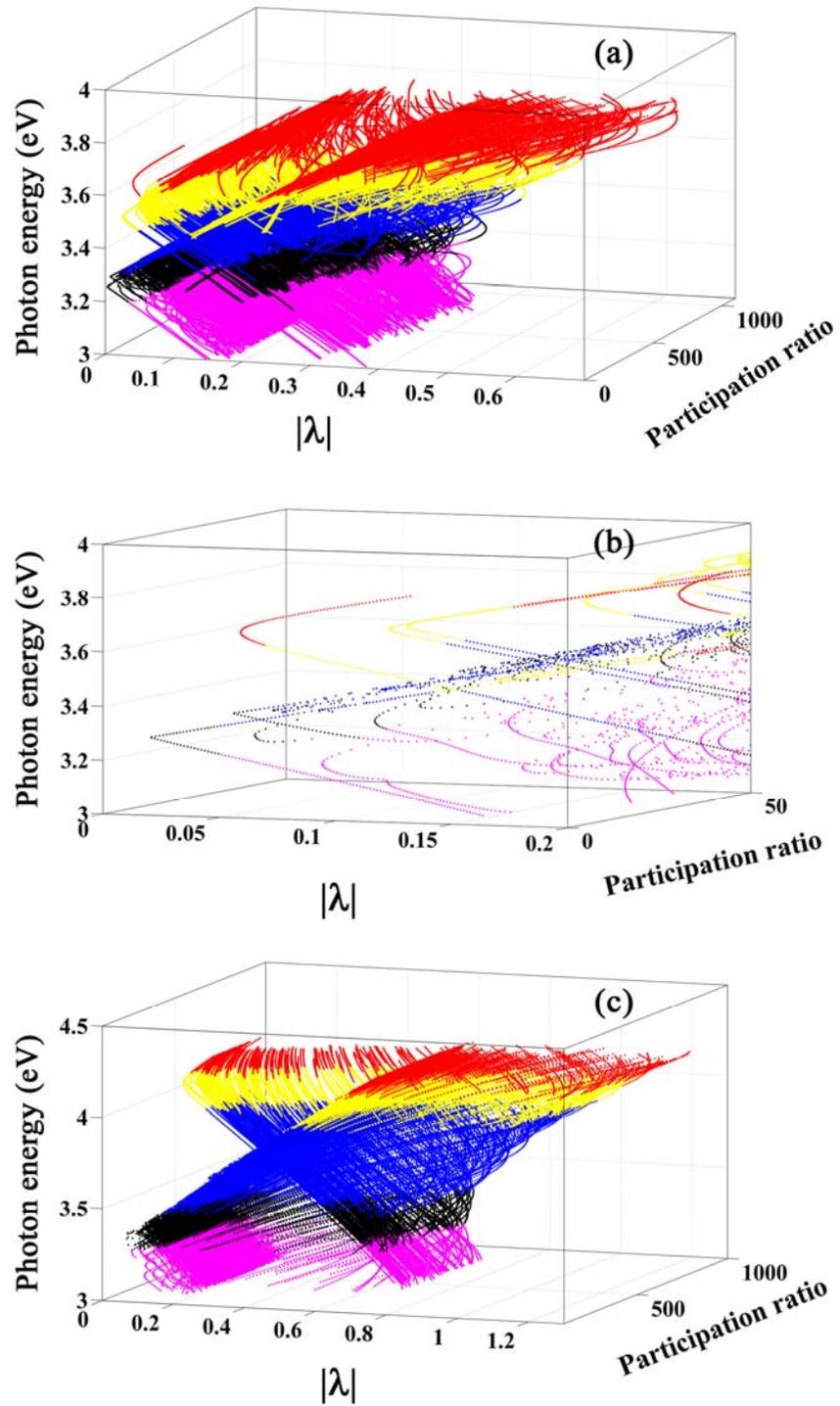

**FIG. 4:** (Color online) 3D trajectory map, defined by the frequency $\omega$ of the modes, the eigenvalue $|\lambda|$ (related to the radiation of the modes), and the participation ratio $P_n$ [related to the eigenvectors (spatial distribution) of the modes], for the (a)

**12-fold QC and (c) square lattice. (b) is the expand view of (a) for small values of $|\lambda|$ and $P_n$. Each plasmonic eigenmode forms a trajectory in 3D parameter space ($\omega$, $|\lambda|$, $P_n$). With the increase of the frequency, $|\lambda|$ will reach a minimum value as the mode is reaching resonance. If such a minimum value tends to zero, it means that the resonant mode is non-radiating. $P_n$ will vary along with the frequency and smaller $P_n$ means the mode is more spatially localized. Different colors mark different frequency regions (same as Fig. 2). The parameters are the same as Fig. 2.**

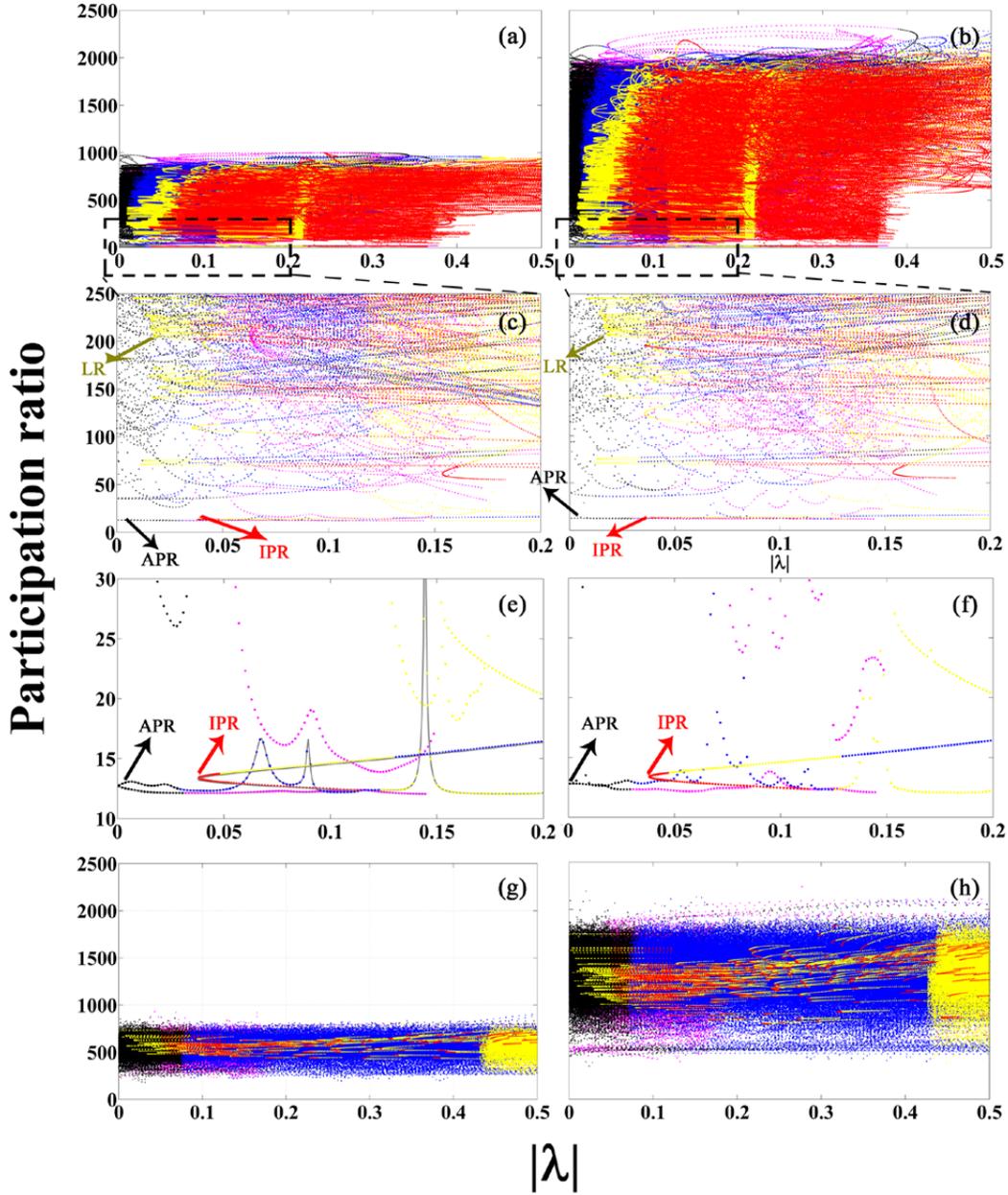

**FIG. 5:** (Color online) (a) The projection of Fig. 4(a) onto the $|\lambda|$ and $P_n$ plane. (b) Projection diagram of a larger-size 12-fold QC with 3529 particles. (c) and (e) [(d) and (f)] are the expanded views of (a) [(b)] with small $|\lambda|$ and small $P_n$. The arrows indicate the resonant points of the APR, IPR, and LR modes. For example, with the help of the grey line that contacts the points in (e), we see that the $|\lambda|$ and $P_n$ of the

**APR mode are the smallest, showing that it is the highest fidelity and highest spatial localization among the eigenmodes. More details are in the text. (g) and (h) are for two square lattices with 1513 and 3613 particles. Different colors mark different frequency regions (same as Fig. 2).**

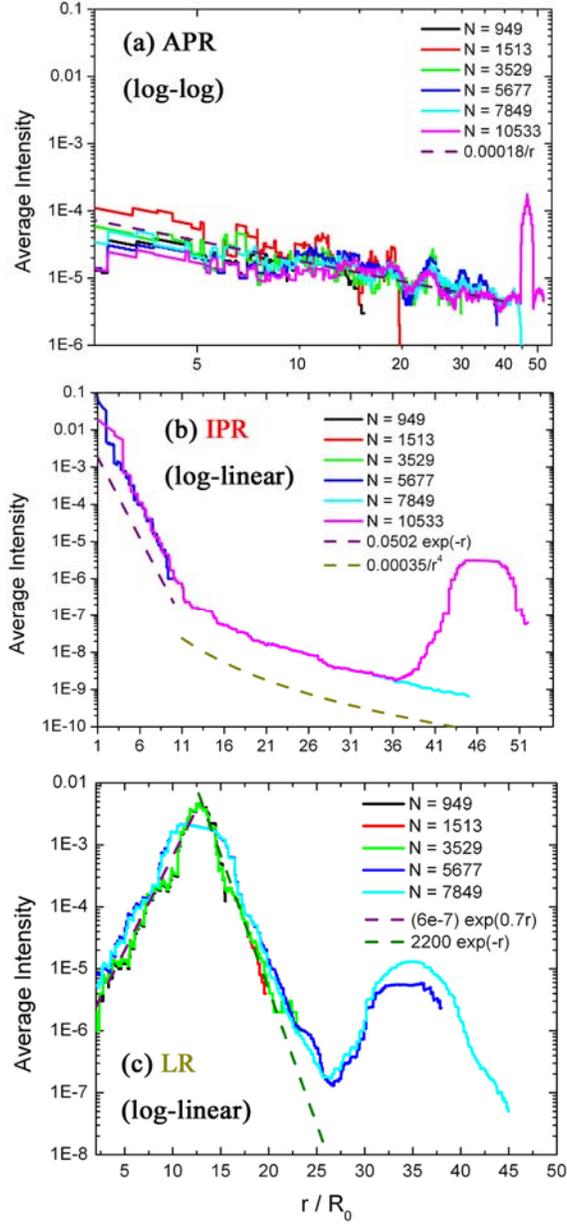

FIG. 6: (Color online) Envelopes of average intensity $I(r)$ for different sizes of QCs when these modes are at resonance. (a) APR mode, log-log scale; (b) IPR mode, log-linear scale; and (c) LR mode, log-linear scale. The dash curves are the fitting curves. Their corresponding mode patterns are shown in the insets of Fig. 3.

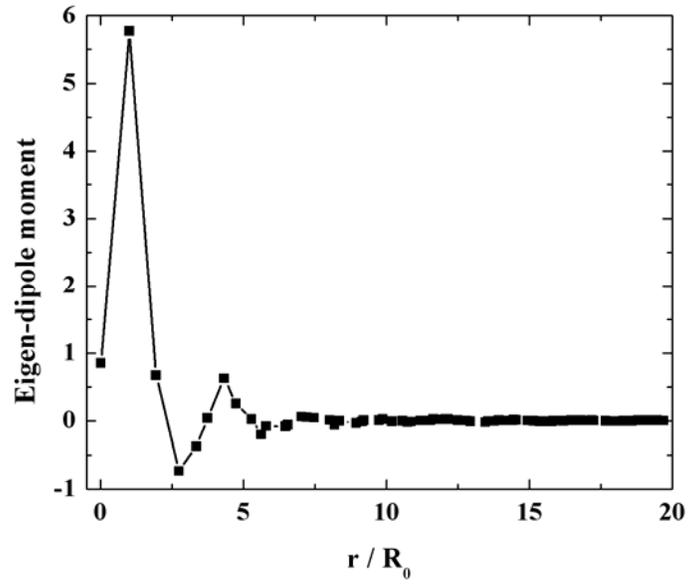

**FIG. 7:** Eigen-dipole moment as a function of $r/R_0$ for the IPR mode at its resonant frequency. The phases of the dipoles oscillate in the first few rings.

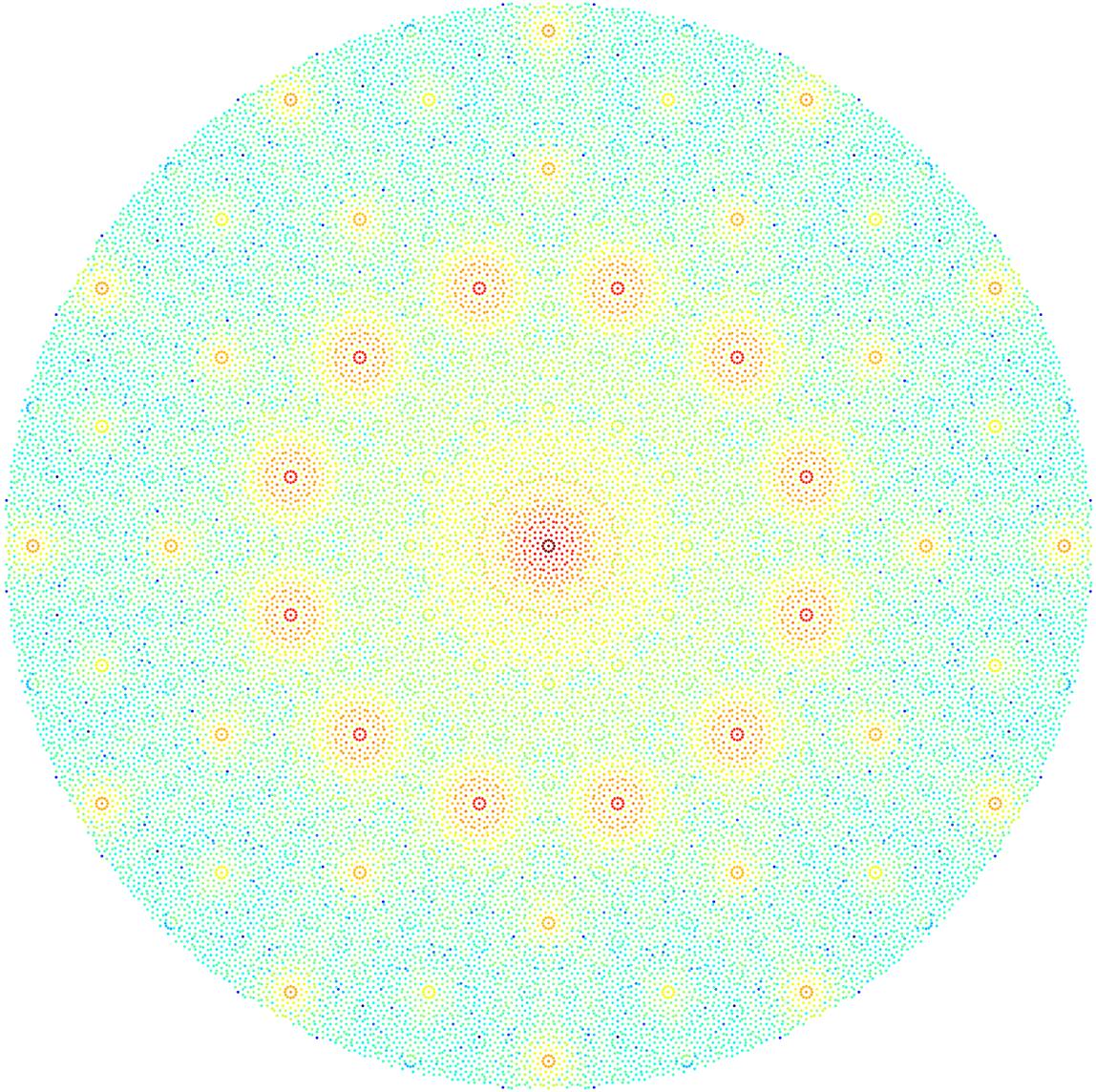

FIG. 8: (Color online) Mode profile (log scale) of the IPR mode of the 12-fold QC with 35269 particles, showing the self-similarity of the mode profile.